\documentclass[aps,prb,reprint,nofootinbib,superscriptaddress,longbibliography]{revtex4-2}

\usepackage{graphicx}
\usepackage{amsmath,amssymb}
\usepackage{bm}
\usepackage{subcaption}
\usepackage{booktabs}
\usepackage{xcolor}
\usepackage[colorlinks=true,citecolor=blue,linkcolor=blue,urlcolor=blue]{hyperref}
\usepackage{caption}
\begin{document}

%============================================================
% TODO (Julian): reemplaza nombre(s), afiliacion(es) y correo
% por los datos reales antes de subir a arXiv.
%============================================================
\title{Structural and Electronic Properties of Bulk $\beta$(2H)-GaSe from First-Principles DFT Calculations with van der Waals Corrections}

\author{Julián A. Aros-González}
\email{jaarosg@udistrital.edu.co}
\affiliation{Programa Académico de Física, Universidad Distrital Francisco José de Caldas, Bogotá, Colombia}

\author{Camilo A. Huertas-Archila}
\affiliation{Programa Académico de Física, Universidad Distrital Francisco José de Caldas, Bogotá, Colombia}

\author{Miguel J. Espitia-Rico}
\affiliation{Programa Académico de Física, Universidad Distrital Francisco José de Caldas, Bogotá, Colombia}

\date{\today}

\begin{abstract}
Gallium selenide (GaSe) is a layered III--VI semiconductor whose bulk crystal serves as the essential energetic reference for modeling the isolated monolayer. We present a systematic density-functional-theory (DFT) study of the centrosymmetric $\beta (\text{2H})$-GaSe polymorph, employing the GGA-PBE functional supplemented with a semiempirical Grimme DFT-D2 dispersion correction to accurately capture the weak van der Waals interlayer coupling. The optimized in-plane lattice parameter agrees well with experimental values (within $\sim 1.3\%$), while the out-of-plane parameter $c$ is overestimated by the DFT-D2 correction relative to experiment, a known limitation of this dispersion scheme when the pairwise $C_6$ coefficients are not specifically fitted for the compound class under study. Electronic structure calculations confirm a non-magnetic, direct-gap profile ($1.12$~eV). Although this magnitude reflects the well-known underestimation of semilocal functionals, the qualitative band topology accurately captures the intrinsic two-dimensional carrier confinement within the bulk material. Consequently, this optimized three-dimensional framework provides a consistent energetic baseline for quantifying exfoliation processes, offering a reliable starting point for future theoretical explorations of the two-dimensional limit, such as surface functionalization for potential spintronic applications.
\end{abstract}

\maketitle

%============================================================
\section{Introduction}
%============================================================

Layered semiconductors held together by weak interlayer van der Waals forces have attracted sustained interest since the isolation of graphene~\cite{Novoselov2004}, motivating the search for two-dimensional (2D) materials with tunable electronic, optical, and mechanical properties~\cite{Mannix2017,Butler2013}. Within the broader family of van der Waals layered compounds, the III--VI metal monochalcogenides, with general formula $MX$ ($M=$ Ga, In; $X=$ S, Se, Te) have emerged as a distinctive class of semiconductors that complements the more widely studied graphene and transition-metal dichalcogenides (TMDs)~\cite{Wu2014,Zhou2018,Huang2016}. Unlike the three-atom-thick X--M--X sandwich structure of TMDs such as MoS$_2$, the III--VI monochalcogenides crystallize in a four-atom-thick X--M--M--X slab with a characteristic metal--metal bond, which qualitatively changes the bonding, the band-gap-versus-thickness trend, and the mechanical response of the material relative to TMDs~\cite{Wu2014,Li2014,Song2023,Chen2020,Huang2016}.

Among the III--VI monochalcogenides, gallium selenide (GaSe) is one of the most extensively studied representatives, owing to its combination of a sizable, tunable electronic gap, a non-centrosymmetric or centrosymmetric crystal structure depending on the polytype, and a pronounced anisotropy inherited from its layered morphology~\cite{Huang2016,Zhou2018}. GaSe has found applications in optoelectronic and photodetection devices~\cite{Sorifi2021,Zou2021}, as one of the most effective materials for nonlinear frequency conversion and terahertz generation via difference-frequency emission~\cite{Yan2017,Serra2025}, and in flexible electronics owing to the low Young's modulus and high mechanical resilience of its layered lattice~\cite{Chen2020,Chuang2018}. GaSe crystallizes in several polymorphs ($\beta$, $\varepsilon$, $\gamma$, $\delta$) that differ only in their interlayer stacking sequence, a consequence of the weak van der Waals coupling between covalently bonded Se--Ga--Ga--Se layers~\cite{AlHattab2024,Yu2024}. Among these, the $\beta$-polymorph (space group $P6_3/mmc$ with a 2H stacking sequence) is the thermodynamically most stable phase and is typically obtained in melt-grown crystals.

An accurate first-principles description of bulk GaSe is not only relevant in its own right, but also serves as the indispensable reference state for quantifying exfoliation, cohesive, and interlayer binding energies in studies of the GaSe monolayer and of GaSe-based van der Waals heterostructures. Because the interlayer bonding in GaSe is governed by dispersive van der Waals interactions that standard semilocal exchange-correlation functionals do not capture, a dispersion correction is required to obtain a reliable interlayer distance and, consequently, a reliable bulk equation of state. In this work, we report a systematic density-functional-theory study of bulk $\beta$-GaSe using the GGA-PBE functional supplemented with the semiempirical Grimme DFT-D2 van der Waals correction. We present a systematic convergence analysis of the numerical parameters that control the accuracy of the plane-wave calculation, the resulting equilibrium structural parameters compared with previous theoretical and experimental work, and the electronic band structure and projected density of states of the optimized bulk crystal.

%============================================================
\section{Computational Methods}
%============================================================

All calculations were performed within the framework of density functional theory (DFT) as implemented in the plane-wave, pseudopotential code \textsc{Quantum ESPRESSO}~\cite{giannozzi2009quantum}. Exchange and correlation effects were described using the generalized-gradient approximation in the Perdew--Burke--Ernzerhof (GGA-PBE) parametrization~\cite{perdew1996generalized}. Because the bulk GaSe structure consists of covalently bonded Se--Ga--Ga--Se layers coupled along the $c$ axis by weak, dispersive interlayer forces, a semiempirical DFT-D2 correction of Grimme~\cite{grimme2006semiempirical} was added to the GGA-PBE total energy in order to describe the van der Waals interlayer bonding. We note that the DFT-D2 correction acts only on the equilibrium geometry and does not modify the underlying GGA-PBE electronic gap.

Core electrons were treated with projector augmented-wave (PAW) pseudopotentials~\cite{Blochl1994,Kresse1999} using the Kresse--Joubert implementation. The explicit valence electron configurations considered were $4s^24p^13d^{10}$ for Ga and $4s^24p^43d^{10}$ for Se. To aid the self-consistent field (SCF) convergence without artificially affecting the semiconductor gap, a Methfessel--Paxton fractional occupation smearing~\cite{methfessel1989high} with a sufficiently small broadening parameter of $0.01$~Ry was applied. The energy and force convergence thresholds for all calculations were strictly set to $10^{-6}$~Ry and $10^{-3}$~Ry/bohr, respectively.

Prior to computing the equilibrium structure and the electronic properties of bulk GaSe, we determined a set of numerical parameters: the plane-wave cutoff energy for the wavefunctions ($E_{\mathrm{cut}}^{\mathrm{wfc}}$), the cutoff energy for the charge density ($E_{\mathrm{cut}}^{\rho}$), and the density of the $\Gamma$-centered Monkhorst--Pack $\mathbf{k}$-point mesh~\cite{monkhorst1976special}. These tests guarantee convergence of the total energy without incurring unnecessary computational cost. The adopted convergence criterion requires that the variation of the total energy between successive values of a given parameter be smaller than $1$~meV per unit cell.

\subsection{Convergence of the wavefunction cutoff energy}

\begin{figure}[t]
    \centering
    \begin{subfigure}[b]{0.48\linewidth}
        \centering
        \includegraphics[width=\textwidth]{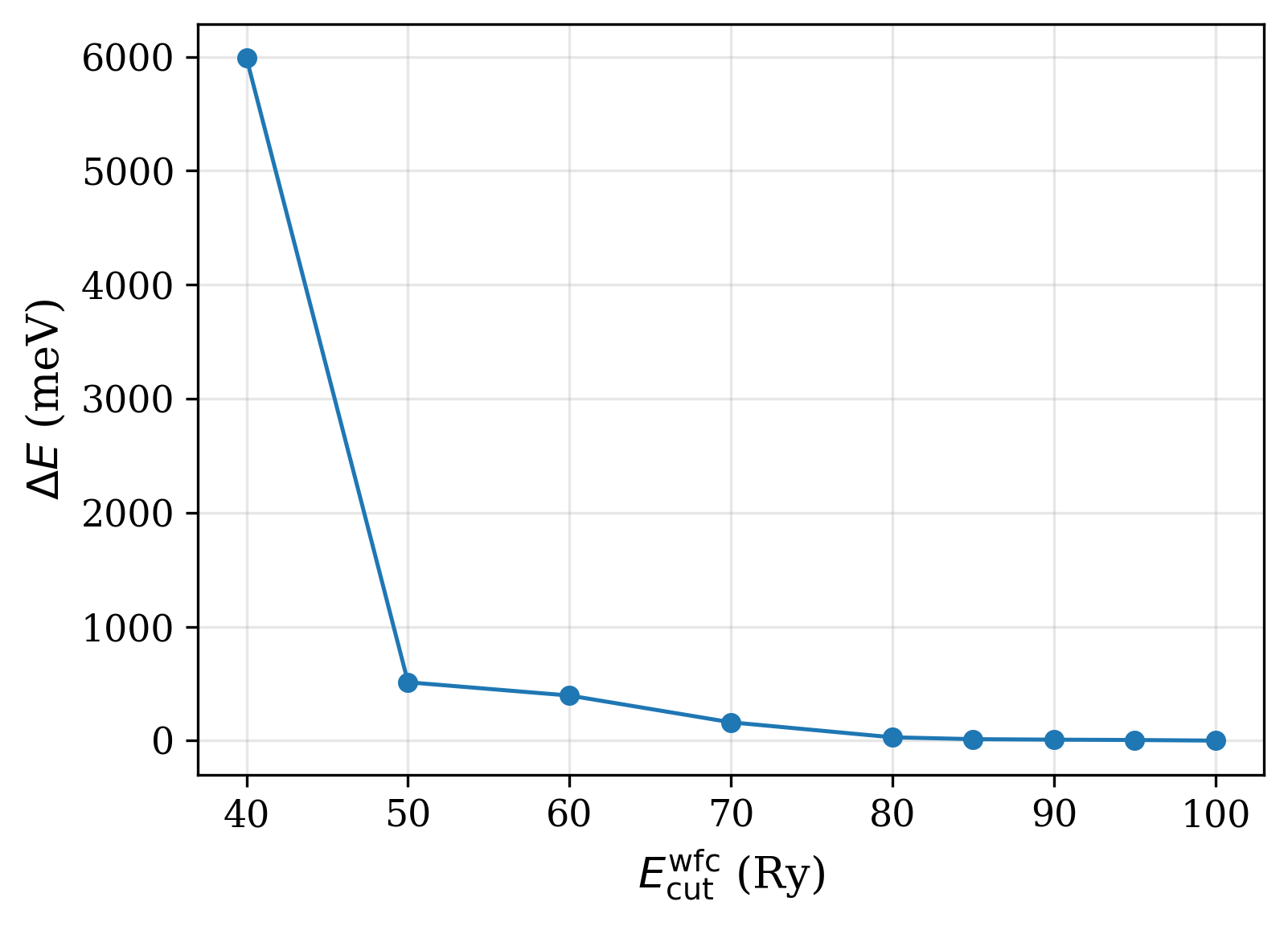}
        \caption{}
        \label{fig:ecutwfc_meV_vol}
    \end{subfigure}
    \hfill
    \begin{subfigure}[b]{0.48\linewidth}
        \centering
        \includegraphics[width=\textwidth]{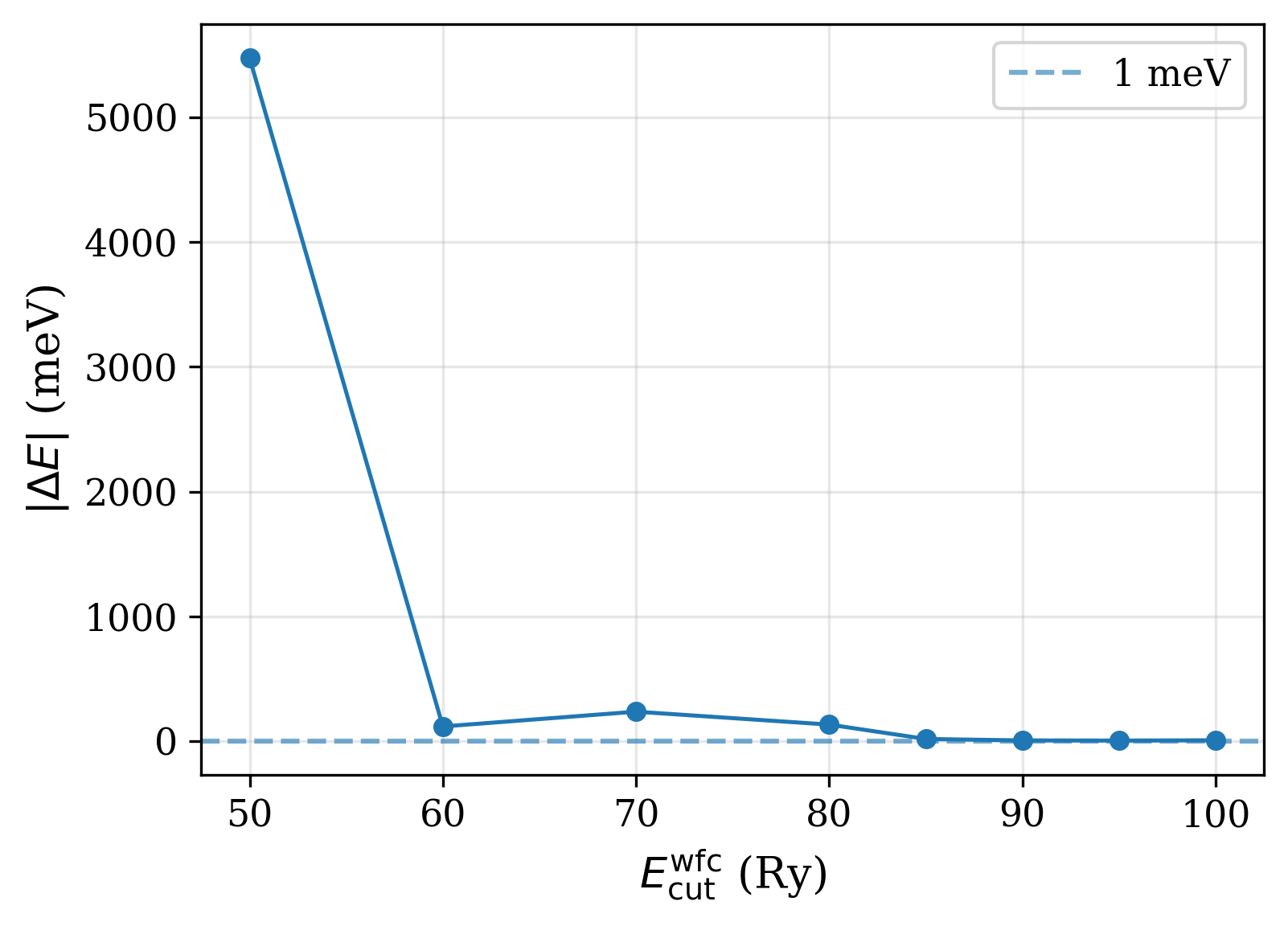}
        \caption{}
        \label{fig:ecutwfc_1meV_vol}
    \end{subfigure}
    \caption{Convergence of the total energy of bulk GaSe with respect to the wavefunction cutoff energy ($E_{\mathrm{cut}}^{\mathrm{wfc}}$): (a) total-energy variation $\Delta E$ relative to the highest cutoff value tested, and (b) absolute value $|\Delta E|$, with the $1$~meV convergence threshold indicated.}
    \label{fig:ecutwfc_vol}
\end{figure}

To determine the optimal size of the plane-wave basis set, the total energy of the system was evaluated for $E_{\mathrm{cut}}^{\mathrm{wfc}}$ values ranging from $40$ to $100$~Ry, keeping other parameters fixed. As shown in Fig.~\ref{fig:ecutwfc_vol}, the total energy exhibits a steep variation at lower cutoffs but rapidly stabilizes. The energy differences fall well below the $1$~meV threshold for cutoffs higher than $85$~Ry. Thus, to ensure a reasonable computational safety margin, a value of $E_{\mathrm{cut}}^{\mathrm{wfc}} = 90$~Ry was selected.

\subsection{Convergence of the charge-density cutoff energy}

\begin{figure}[t]
    \centering
    \begin{subfigure}[b]{0.48\linewidth}
        \centering
        \includegraphics[width=\textwidth]{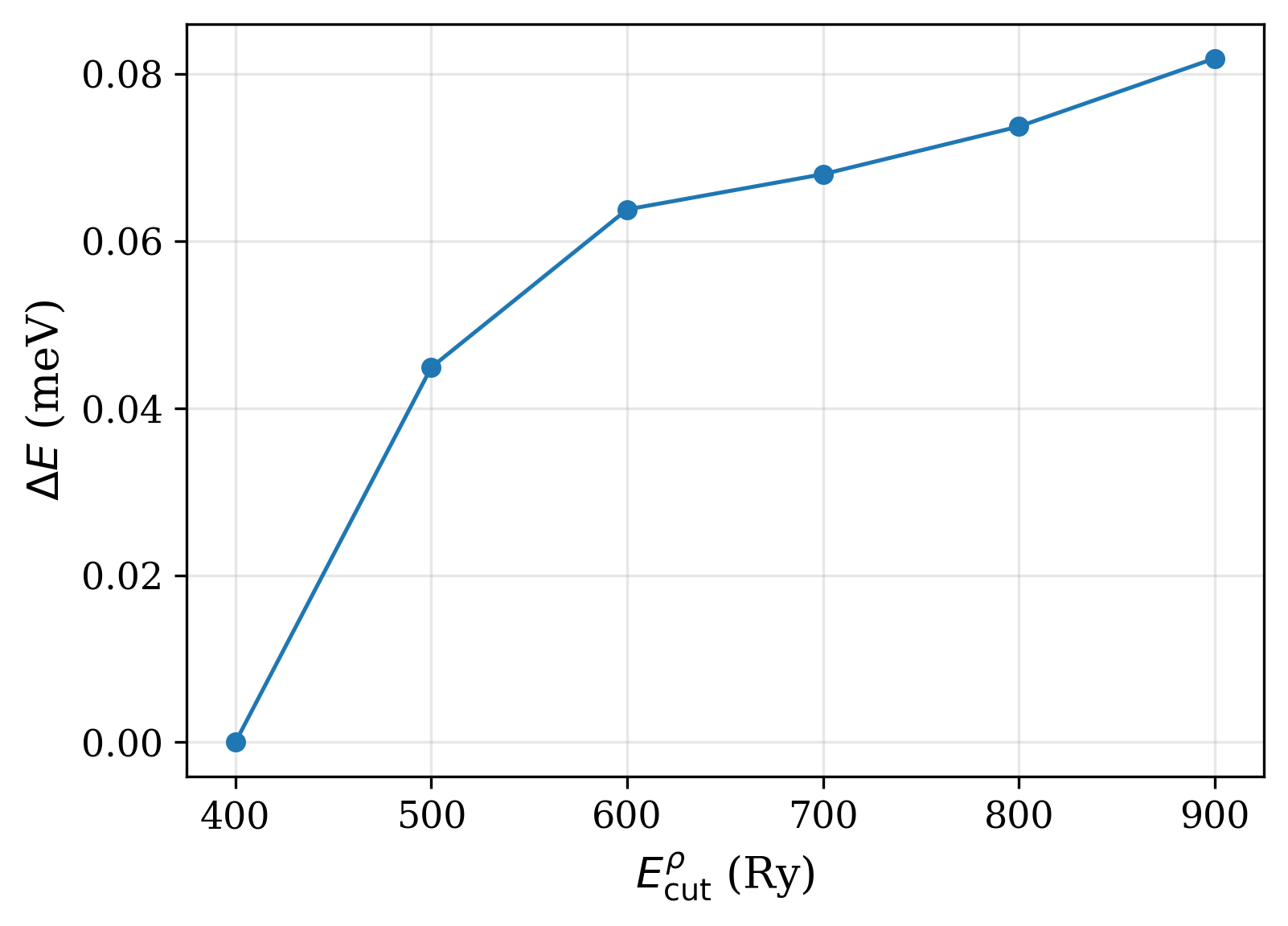}
        \caption{}
        \label{fig:ecutrho_meV_vol}
    \end{subfigure}
    \hfill
    \begin{subfigure}[b]{0.48\linewidth}
        \centering
        \includegraphics[width=\textwidth]{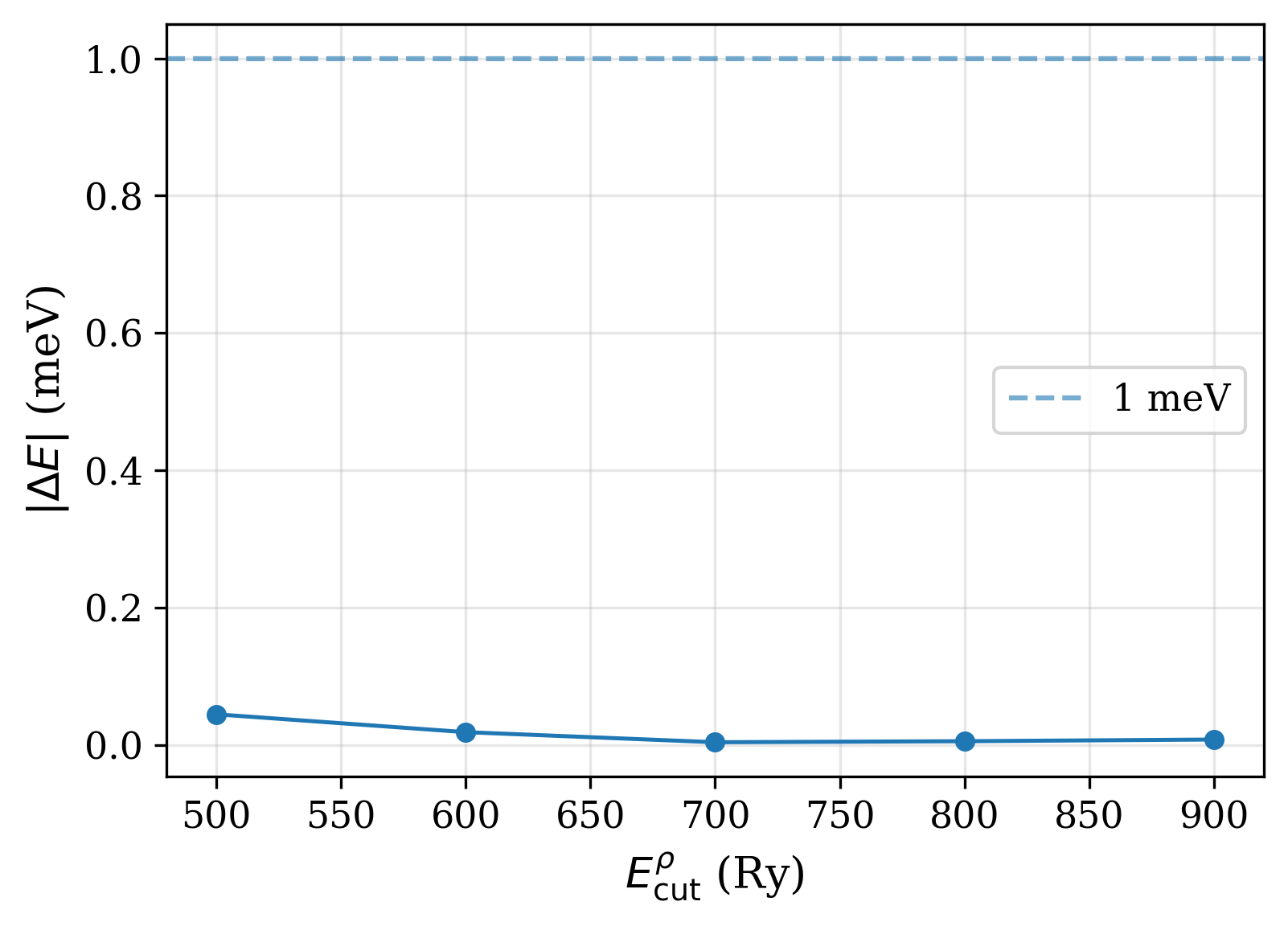}
        \caption{}
        \label{fig:ecutrho_1meV_vol}
    \end{subfigure}
    \caption{Convergence of the total energy of bulk GaSe with respect to the charge-density cutoff energy ($E_{\mathrm{cut}}^{\rho}$): (a) total-energy variation $\Delta E$ relative to the highest cutoff value tested, and (b) absolute value $|\Delta E|$, with the $1$~meV convergence threshold indicated.}
    \label{fig:ecutrho_vol}
\end{figure}

The cutoff energy for the charge density determines the representation of the electronic density in reciprocal space and must be significantly larger than $E_{\mathrm{cut}}^{\mathrm{wfc}}$ when using PAW pseudopotentials. For $E_{\mathrm{cut}}^{\rho}$, explored within the $400$--$900$~Ry range, the successive energy variations were consistently smaller than $0.1$~meV (Fig.~\ref{fig:ecutrho_vol}). Following the typical $E_{\mathrm{cut}}^{\rho}/E_{\mathrm{cut}}^{\mathrm{wfc}}$ ratio for this type of pseudopotential, we adopted $E_{\mathrm{cut}}^{\rho} = 700$~Ry.

\subsection{Convergence of the Brillouin-zone sampling}

\begin{figure}[t]
    \centering
    \begin{subfigure}[b]{0.48\linewidth}
        \centering
        \includegraphics[width=\textwidth]{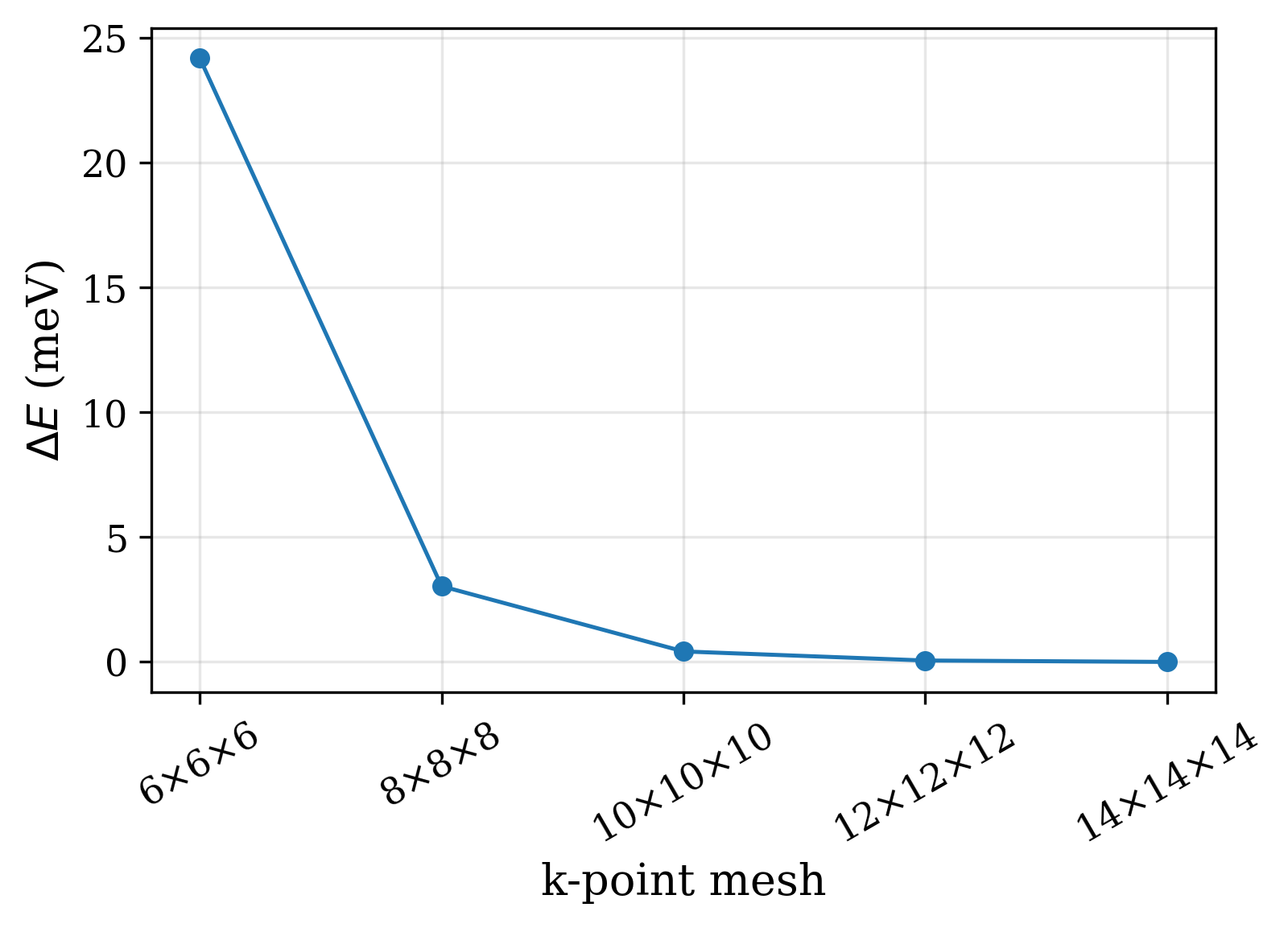}
        \caption{}
        \label{fig:kpoint_meV_vol}
    \end{subfigure}
    \hfill
    \begin{subfigure}[b]{0.48\linewidth}
        \centering
        \includegraphics[width=\textwidth]{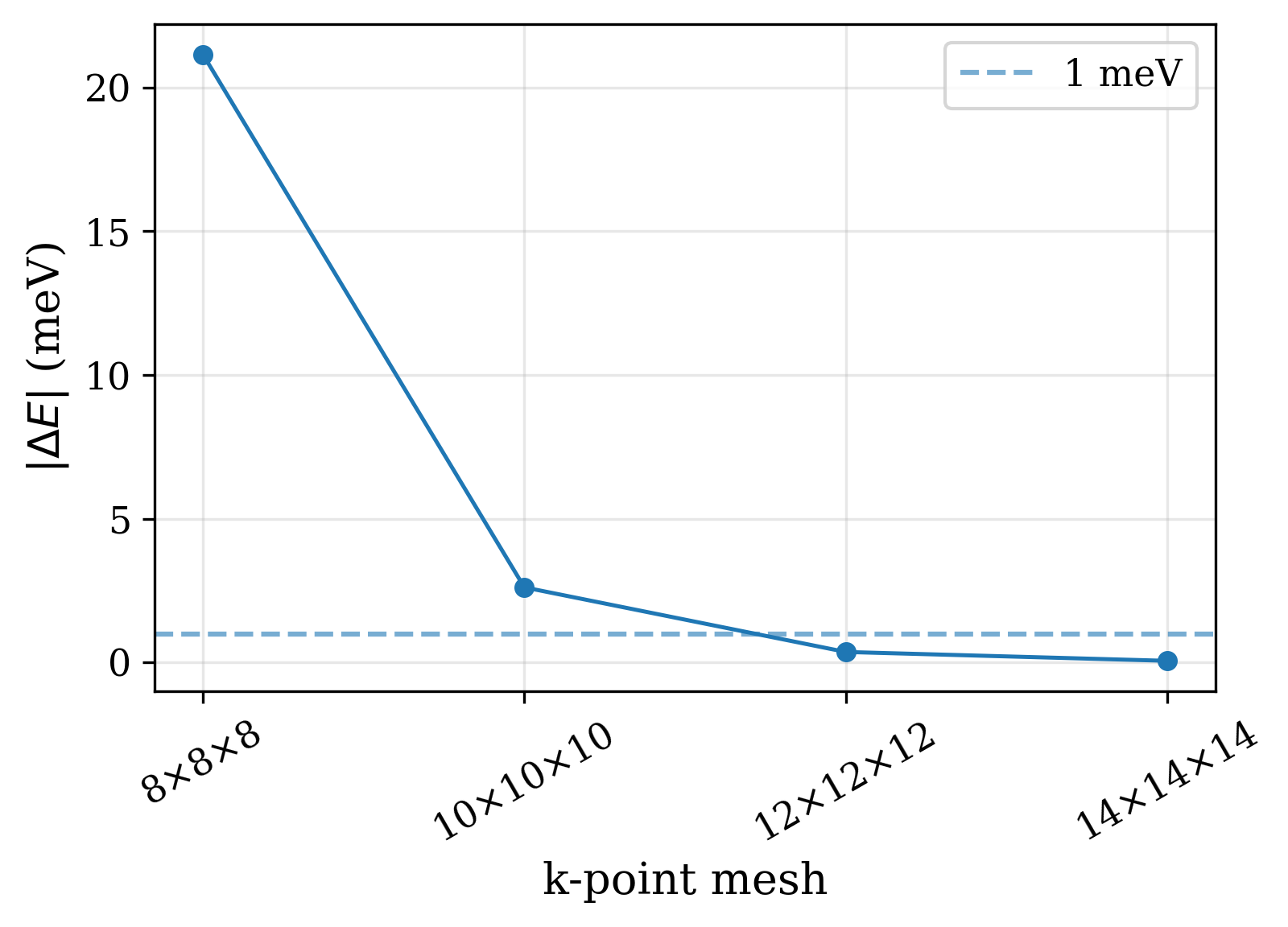}
        \caption{}
        \label{fig:kpoint_1meV_vol}
    \end{subfigure}
    \caption{Convergence of the total energy of bulk GaSe with respect to the density of the Monkhorst--Pack $\mathbf{k}$-point mesh: (a) total-energy variation $\Delta E$ relative to the densest mesh tested, and (b) absolute value $|\Delta E|$, with the $1$~meV convergence threshold indicated.}
    \label{fig:kpoint_vol}
\end{figure}

The sampling of the Brillouin zone was tested using Monkhorst--Pack meshes ranging from $6\times6\times6$ to $14\times14\times14$. The variation in total energy drops below the $1$~meV threshold when transitioning from $10\times10\times10$ to $12\times12\times12$, and stabilizes under $0.2$~meV for denser meshes (Fig.~\ref{fig:kpoint_vol}). Therefore, the $12\times12\times12$ mesh was established as the optimal reference.

The converged computational parameters are summarized in Table~\ref{tab:convergencia_vol}. With these parameters fixed, the geometric optimization of the bulk phase was performed. Rather than relying on a standard variable-cell relaxation algorithm, the equilibrium lattice parameters were determined through a systematic manual scan to eliminate potential computational errors derived from Pulay stress. At each scanned value, the atomic positions were relaxed at fixed cell shape and volume (constant-cell ionic relaxation). First, the in-plane parameter $a$ was optimized by minimizing the resulting total energy at a fixed $c/a$
ratio. Subsequently, the out-of-plane parameter $c$ was refined in the same manner while keeping $a$ fixed at its optimized value. The energy minima were analytically extracted using quadratic fits (as detailed in Sec.~\ref{sec:structure}), ensuring a rigorous description of the van der Waals-governed interlayer spacing.

\begin{table}[t]
\centering
\caption{Converged computational parameters for bulk GaSe. The values of $a$ and $c$ correspond to the energy minimum obtained from a second-order polynomial fit (Sec.~\ref{sec:structure}).}
\label{tab:convergencia_vol}
\begin{tabular}{ll}
\toprule
\textbf{Parameter} & \textbf{Optimal value} \\
\midrule
$E_{\text{cut}}^{\text{wfc}}$ & 90~Ry \\
$E_{\text{cut}}^{\rho}$ & 700~Ry \\
$\mathbf{k}$-point mesh & $12\times12\times12$ (Monkhorst--Pack) \\
Lattice parameter $a$ & 7.1891~bohr (3.804~\AA) \\
Lattice parameter $c$ & 31.9802~bohr (16.923~\AA) \\
Ratio $c/a$ & 4.448 \\
\bottomrule
\end{tabular}
\end{table}
%============================================================
\section{Results and Discussion}
%============================================================

\subsection{Crystal structure and lattice-parameter optimization}
\label{sec:structure}

\begin{figure}[t]
    \centering
    \includegraphics[width=0.3\linewidth]{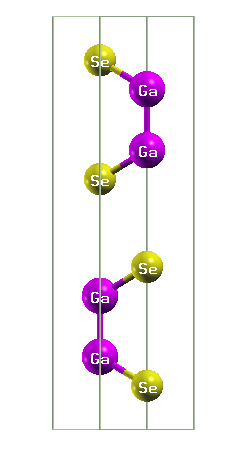}
    \caption{Unit cell of bulk GaSe in the $\beta$(2H)-polymorph [$P6_3/mmc$ (No.~194)], showing the Se--Ga--Ga--Se layer stacking along the $c$ axis.}
    \label{fig:bulk_unitaria}
\end{figure}

Bulk GaSe crystallizes in the $\beta$(2H)-polymorph, with space group $P6_3/mmc$ (No.~194) and hexagonal symmetry. The unit cell contains eight atoms arranged in layers stacked along the $c$ axis, following the characteristic Se--Ga--Ga--Se sequence (Fig.~\ref{fig:bulk_unitaria}). Within each layer, Ga atoms form covalent Ga--Ga bonds directed along the $c$ axis, while Se atoms occupy the outer faces, covalently bonded to the adjacent Ga atoms; successive layers are coupled by low-energy van der Waals interactions, which give the material its pronounced anisotropy and its ease of mechanical exfoliation, both hallmarks of the III--VI layered family. The $\beta$-phase is the highest-symmetry polytype of GaSe, with 24 symmetry operations and a site multiplicity of 4; successive layers are related by a translation combined with a $60^{\circ}$ rotation.

\begin{figure}[t]
    \centering
    \includegraphics[width=0.85\linewidth]{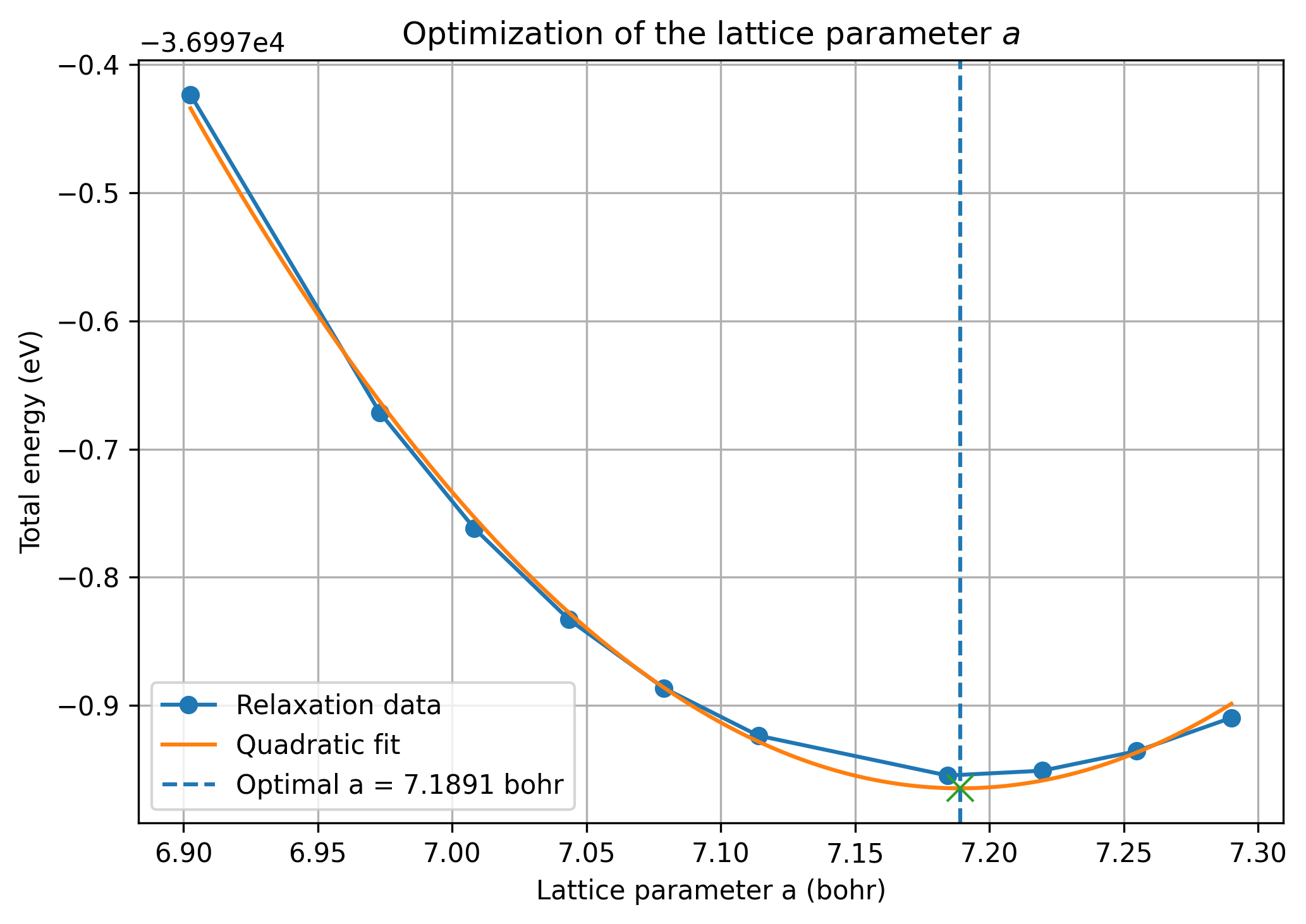}
    \caption{Total energy of bulk GaSe as a function of the lattice parameter $a$, obtained from constant-cell ionic relaxations over the interval $6.90$--$7.30$~bohr.}
    \label{fig:a_vol}
\end{figure}

With the converged numerical parameters of table~\ref{tab:convergencia_vol}, the equilibrium lattice parameter $a$ was determined by performing constant-cell ionic relaxations (fixed lattice, relaxed atomic positions) for values of $a$ between $6.90$ and $7.30$~bohr, keeping the ratio $c/a$ fixed, and fitting the resulting total energy to a second-order polynomial,
\begin{equation}
E(a) = 6.4665\,a^{2} - 92.9761\,a - 36663.76 \quad [\mathrm{eV}],
\label{eq:Ea}
\end{equation}
whose minimum, obtained analytically, gives $a = 7.1891$~bohr ($3.804$~\AA) [Fig.~\ref{fig:a_vol}]. The positive curvature of the fitted polynomial confirms a genuine minimum, and the fit reproduces the relaxation data closely around the equilibrium point.

\begin{figure}[t]
    \centering
    \includegraphics[width=0.85\linewidth]{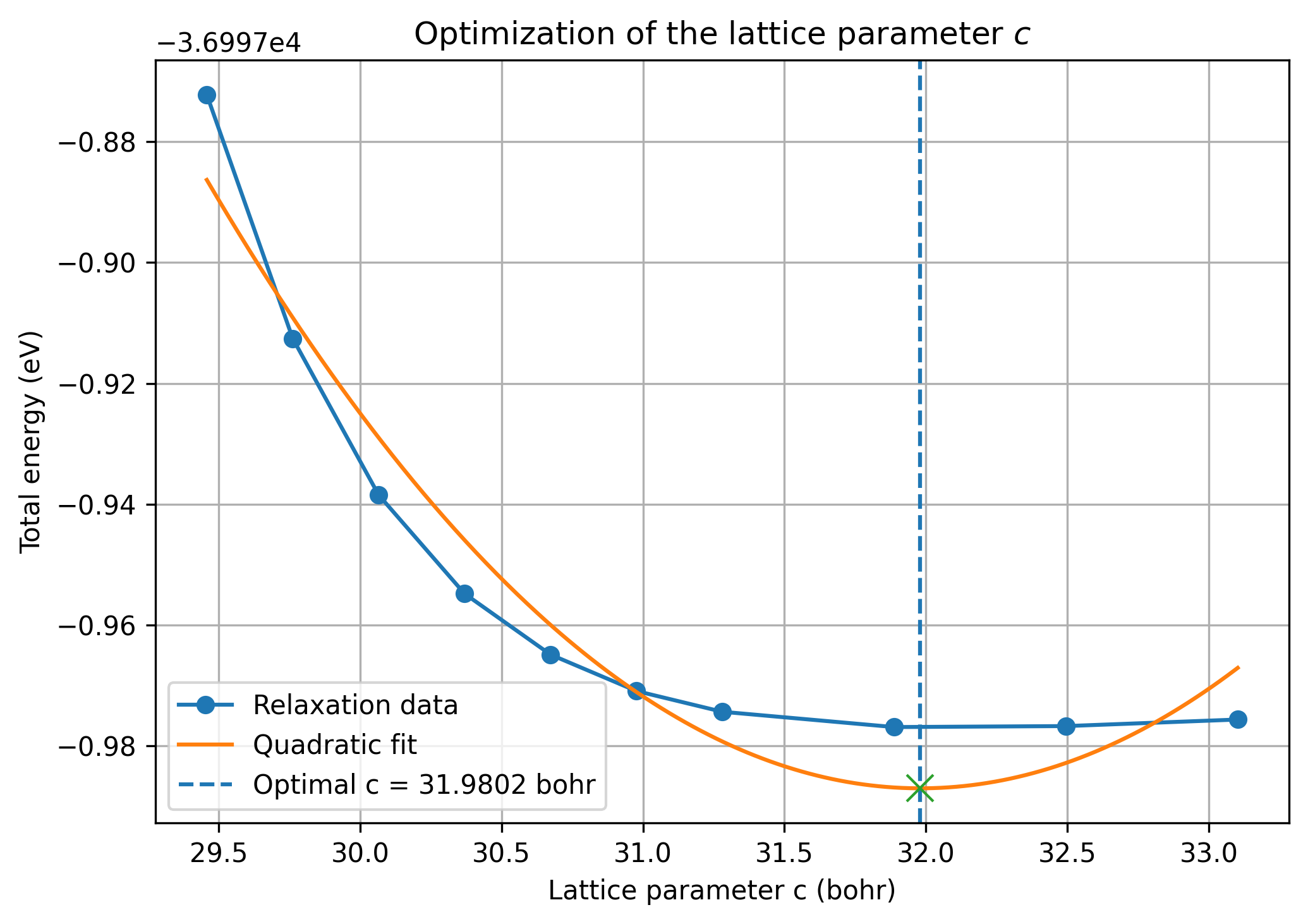}
    \caption{Total energy of bulk GaSe as a function of the lattice parameter $c$, obtained from constant-cell ionic relaxations over the interval $29.5$--$33.0$~bohr.}
    \label{fig:c_vol}
\end{figure}

With $a$ fixed at its optimal value, the lattice parameter $c$ was optimized over the interval $29.5$--$33.0$~bohr using the same constant-cell ionic-relaxation procedure, and the resulting total energy fitted to
\begin{equation}
E(c) = 0.015844\,c^{2} - 1.013387\,c - 36981.78 \quad [\mathrm{eV}],
\label{eq:Ec}
\end{equation}
whose minimum gives $c = 31.9802$~bohr ($16.923$~\AA) [Fig.~\ref{fig:c_vol}]. For $c$ larger than approximately $31.5$~bohr the calculated energy becomes nearly flat, with a slight upward trend, while the quadratic fit more accurately captures the central region of the curve. The resulting aspect ratio, $c/a \approx 4.45$, is consistent with values reported in the literature for this structure.

Table~\ref{tab:parametros_red_bulk} places the obtained lattice parameters in the context of previous theoretical and experimental work. For the in-plane parameter $a$, all theoretical calculations (irrespective of the functional or method employed) systematically overestimate the experimental value, with deviations ranging from $0.6\%$~\cite{Rak2009GaSeDefects} to $2.1\%$~\cite{Ghalouci2013GaSe}; the present work yields an intermediate deviation of $\sim 1.3\%$. This trend is inherent to the GGA-PBE functional and reflects its known underestimation of intralayer bonding forces. For the out-of-plane parameter $c$, the spread among references is markedly larger, highlighting the general difficulty of accurately describing the van der Waals-governed interlayer distance. The three pairs of values reported in Ref.~\cite{Rak2009GaSeDefects} correspond to different pseudopotentials and approximations within the same study, which explains their non-monotonic variation between $a$ and $c$. The value of $c$ obtained in this work ($16.923$~\AA) is the largest in the series, exceeding both the theoretical result of Ref.~\cite{Ghalouci2013GaSe} ($16.393$~\AA) and the highest experimental reference, Ref.~\cite{DeBlasi1989GaChalcogenides} ($16.10$~\AA). This behavior contrasts with the overbinding tendency typically reported for DFT-D2 in layered van der Waals systems\footnote{In graphite, for instance, DFT-D2 is well known to \emph{underestimate} the interlayer spacing (yielding $\sim 3.21$~\AA\ against an experimental value of $3.34$~\AA)~\cite{VanTroeye2016}.}, i.e., the correction there overestimates the interlayer attraction rather than underestimates it. The overestimated $c$ obtained here for GaSe therefore suggests that the standard DFT-D2 parametrization instead \emph{underestimates} the net interlayer attraction in this system, likely reflecting the generic, element-pairwise nature of the Grimme~\cite{grimme2006semiempirical} $C_6$ coefficients, which were not fitted specifically for III--VI layered compounds. Consistent with this, the obtained ratio $c/a = 4.448$ is somewhat higher than the literature range ($4.25$--$4.28$), reflecting the overestimated interlayer spacing discussed above; the overall hexagonal cell geometry is nonetheless reproduced satisfactorily at the qualitative level.

\begin{table*}[t]
\centering
\caption{Calculated lattice parameters for $\beta$-GaSe [$P6_3/mmc$ (No.~194)] and previously reported experimental and theoretical data.}
\label{tab:parametros_red_bulk}
\begin{tabular*}{\textwidth}{@{\extracolsep{\fill}}lccc}
\toprule
\textbf{Reference} & \textbf{$a_0$ (\AA)} & \textbf{$c_0$ (\AA)} & \textbf{$c/a$} \\
\midrule
This work & 3.804 & 16.923 & 4.448 \\
Theory~\cite{Ghalouci2013GaSe}  & 3.830 & 16.393 & 4.28 \\
Theory~\cite{Rak2009GaSeDefects} & 3.77--3.80--3.83 & 16.39--16.17--16.29 & --- \\
Experiment~\cite{Whitehouse1978GaSeGrowth} & $3.755 \pm 0.002$ & $15.955 \pm 0.012$ & --- \\
Experiment~\cite{Kuhn1975GaSeStructure} & 3.752--3.755 & 15.95--15.94 & --- \\
Experiment~\cite{DeBlasi1989GaChalcogenides} & $3.74 \pm 0.01$ & $16.10 \pm 0.05$ & --- \\
\bottomrule
\end{tabular*}
\end{table*}

\subsection{Exfoliation and interlayer binding energies}
\label{sec:exfoliation}

To establish quantitatively that bulk GaSe is a viable precursor for the isolated monolayer, we computed two independent, complementary energies: the exfoliation energy from a slab-convergence method, and the interlayer binding energy evaluated directly from the bulk unit cell.

The exfoliation energy required to remove a single layer from an $N$-layer freestanding slab is
\begin{equation}
E_{\mathrm{exf}} = -\frac{E_{N} - E_{N-1} - E_{1}}{A_{\mathrm{sup}}},
\label{eq:Eexf}
\end{equation}
where $E_N$ and $E_{N-1}$ are the total energies of slabs with $N$ and $N-1$ layers, $E_1$ is the total energy of the isolated monolayer, and $A_{\mathrm{sup}}$ is the basal area of the surface unit cell. Using slabs of $N=4$ and $N=3$ layers we obtain
\begin{equation}
E_{\mathrm{exf}} = 0.002079~\mathrm{Ry/\AA^2} = 28.295~\mathrm{meV/\AA^2}.
\end{equation}

As an independent check, we evaluated the interlayer binding energy directly from the fully relaxed bulk unit cell, which contains $N_{\mathrm{cell}}=2$ layers (Sec.~\ref{sec:structure}),
\begin{equation}
E_{b} = -\frac{E_{\mathrm{bulk}}/N_{\mathrm{cell}} - E_{1}}{A_{\mathrm{sup}}},
\label{eq:Eb}
\end{equation}
which gives
\begin{equation}
E_{b} = 0.002046~\mathrm{Ry/\AA^2} = 27.850~\mathrm{meV/\AA^2}.
\end{equation}

The two independent estimates agree to within $1.6\%$, cross-validating the two computational routes. This agreement is consistent with, but does not by itself establish, convergence of the exfoliation energy with respect to slab thickness; a dedicated test across additional values of $N$ would be required to confirm this point directly. Both values fall in the range of tens of meV/\AA$^2$ characteristic of easily exfoliable van der Waals layered materials~\cite{bjorkman2012van}, comparable to values reported for related III--VI monochalcogenide monolayers such as GaS and GeSe~\cite{Querne2023}. This indicates that mechanical exfoliation of bulk $\beta$-GaSe into isolated Se--Ga--Ga--Se monolayers is energetically favorable, and supports the use of the bulk lattice optimized in Sec.~\ref{sec:structure} as a consistent energetic reference for the monolayer.

\subsection{Electronic properties}

\begin{figure}[h]
    \centering
    \includegraphics[width=\linewidth]{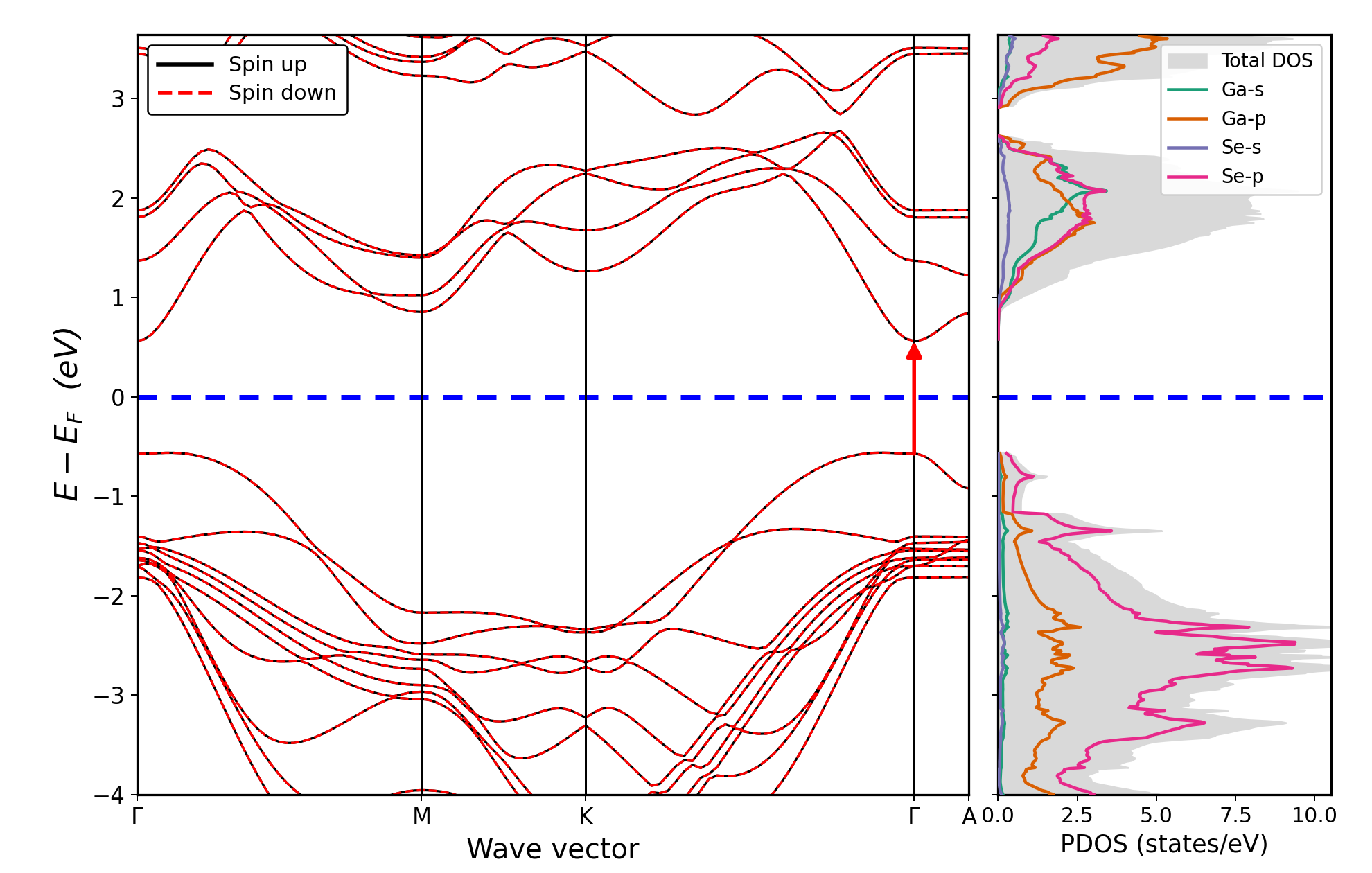}
    \caption{Electronic band structure and projected density of states (PDOS) of bulk GaSe. The Fermi level is set to zero (dashed blue line).}
    \label{fig:bandasvolumen}
\end{figure}

The electronic band structure of the optimized bulk GaSe crystal, together with the projected density of states (PDOS), is shown in Fig.~\ref{fig:bandasvolumen}. Bulk GaSe is found to be a non-magnetic semiconductor, as is also the case for the GaSe monolayer~\cite{Cao2015,Wu2014}, with a direct gap at the $\Gamma$ point of $1.12$~eV. The spin-polarized calculation confirms this: the spin-up and spin-down bands (solid black and dashed red lines, respectively) are perfectly degenerate throughout the Brillouin zone, yielding a total magnetic moment of zero.

The calculated direct gap of $1.12$~eV underestimates the experimental room-temperature optical bandgap of bulk GaSe, typically reported around $2.0$~eV~\cite{Norkus2020}, with photoluminescence measurements pinpointing the fundamental emission at $2.005$~eV~\cite{Chuang2018}. Such quantitative underestimation is an expected artifact of semilocal exchange-correlation functionals like GGA-PBE, stemming from the self-interaction error and the lack of a derivative discontinuity in the potential~\cite{Capelle2006DFT,martin2004electronic}. Despite this numerical deviation, the functional provides a well-defined qualitative description of the electronic structure, correctly capturing the direct ($\Gamma$-$\Gamma$) nature of the optical transition, the orbital hybridization character, and the band dispersion topology.
The band structure is plotted along the high-symmetry path of the hexagonal Brillouin zone $\Gamma \rightarrow M \rightarrow K \rightarrow \Gamma \rightarrow A$. Along the $c$-axis direction ($\Gamma \rightarrow A$), the band dispersion is very flat, reflecting a large effective mass and a nearly two-dimensional electronic confinement even in the bulk phase~\cite{Norkus2020,Eremeev2020}. In the basal plane (direction $\Gamma \rightarrow M \rightarrow K$), the bands display a more pronounced, parabolic dispersion, indicating high in-plane carrier mobility within the covalently bonded Se--Ga--Ga--Se planes~\cite{Norkus2020}.

The projected density of states shows that the valence band has orbital contributions from both Ga-$p$ and Se-$p$ states. The conduction band near the Fermi level, on the other hand, arises from a hybridization of Ga and Se $p$ states together with a Ga-$s$ contribution.

%============================================================
\section{Conclusions}
%============================================================

We established the equilibrium structural and electronic properties of the $\beta$-polymorph of GaSe through a DFT framework incorporating the Grimme DFT-D2 dispersion correction. The optimized lattice parameters ($a = 3.804$~\AA, $c = 16.923$~\AA) show that while standard semilocal exchange-correlation functionals fail to capture weak interlayer forces, the semiempirical van der Waals correction stabilizes the layered geometry and reproduces the in-plane parameter in good agreement with experiment. The DFT-D2 scheme, however, overestimates the out-of-plane parameter $c$ and, correspondingly, the $c/a$ aspect ratio relative to experimental benchmarks, pointing to an underestimation of the net interlayer attraction by the generic $C_6$ coefficients used for Ga and Se; the overall hexagonal cell geometry is nonetheless captured at the qualitative level.

The calculated electronic band structure corroborates the non-magnetic, direct-gap semiconducting nature of the bulk crystal. The pronounced contrast in band dispersion (exhibiting remarkably flat profiles along the $c$-axis and parabolic curves in the basal plane) highlights the intrinsic two-dimensional electronic confinement that persists even within the bulk crystal.

The pronounced structural and electronic anisotropy demonstrated by our results naturally highlights the viability of isolating the two-dimensional GaSe monolayer. Consequently, the optimized bulk lattice presented here provides the essential energetic reference required to accurately quantify exfoliation and cohesive energies. Furthermore, given the intrinsically non-magnetic nature of pristine GaSe, isolating its single-layer form exposes covalently bonded selenium surfaces that are highly amenable to functionalization. This robust three-dimensional baseline therefore offers a reliable starting point for future theoretical explorations---such as the adsorption of $3d$ transition metals---aimed at inducing tailored magnetic properties in the two-dimensional limit for potential spintronic applications.

\newpage
%============================================================
\begin{acknowledgments}
The authors acknowledge the computational resources provided by the Faculty of Mathematics and Natural Sciences at Francisco José de Caldas District University.
\end{acknowledgments}
%============================================================

\bibliography{referencias}

\end{document}